\documentclass[runningheads]{llncs}
\usepackage[utf8]{inputenc}
\usepackage{listings}
\usepackage{color}
\usepackage{algorithm2e}
\makeatletter
\renewcommand{\@algocf@capt@plain}{above}
\makeatother
\usepackage{caption}
\usepackage{amsmath}
\usepackage{amssymb}
\usepackage{xspace}
\usepackage[colorinlistoftodos]{todonotes}

\newcommand{\vct}[1]{\ensuremath{\boldsymbol{#1}}}
\newcommand{\mat}[1]{\ensuremath{\mathbf{#1}}}

\newcommand{\T}{\ensuremath{\top}}

\newcommand{\argmin}{\operatornamewithlimits{\arg\,\min}}

\newcommand{\myparagraph}[1]{\smallskip \noindent \textbf{#1.}}
\newcommand{\ie}{{i.e.}\xspace}

\newcommand{\refToSection}[1]{Section~\ref{sec:#1}\xspace}

\title{Explaining Vulnerabilities of Deep Learning to Adversarial Malware Binaries}
\author{
    Luca Demetrio\inst{1} 
    \and Battista Biggio\inst{2,3} 
    \and Giovanni Lagorio\inst{1} 
    \and Fabio Roli\inst{2,3}
    \and Alessandro Armando\inst{1}
}

\institute{
    CSecLab, University of Genova,
    Genova, Italy\\
    \email{luca.demetrio@dibris.unige.it,\\\{giovanni.lagorio, alessandro.armando\}@unige.it}
    \and
    PRALab, Department of Electrical and Electronic Engineering,\\
    University of Cagliari, Cagliari, Italy\\
 	\email{\{battista.biggio, fabio.roli\}@diee.unica.it}\\
 	\and Pluribus One, Italy
}

\authorrunning{L. Demetrio, B. Biggio, G. Lagorio, F. Roli and A. Armando}
\titlerunning{Explaining Vulnerabilities of Deep Learning to Adversarial Malware Binaries}

\begin{document}

\maketitle

\begin{abstract}
Recent work has shown that deep-learning algorithms for malware detection are also susceptible to adversarial examples, i.e., carefully-crafted perturbations to input malware that enable misleading classification. 
Although this has questioned their suitability for this task, it is not yet clear why such algorithms are easily fooled also in this particular application domain.
In this work, we take a first step to tackle this issue by leveraging explainable machine-learning algorithms developed to interpret the black-box decisions of deep neural networks. 
In particular, we use an explainable technique known as feature attribution to identify the most influential input features contributing to each decision, and adapt it to provide meaningful explanations to the classification of malware binaries. 
In this case, we find that a recently-proposed convolutional neural network does not learn any meaningful characteristic for malware detection from the data and text sections of executable files, but rather tends to learn to discriminate between benign and malware samples based on the characteristics found in the file header. 
Based on this finding, we propose a novel attack algorithm that generates adversarial malware binaries by only changing few tens of bytes in the file header.
With respect to the other state-of-the-art attack algorithms, our attack does not require injecting any padding bytes at the end of the file, and it is much more efficient, as it requires manipulating much fewer bytes.
\end{abstract}

\section{Introduction}
\label{sec:intro}

Despite their impressive performance in many different tasks, deep-learning algorithms have been shown to be easily fooled by \emph{adversarial examples}, \ie, carefully-crafted perturbations of the input data that cause misclassifications~\cite{biggio2013evasion,szegedy14-iclr,goodfellow6572explaining,papernot2016limitations,carlini2017towards,biggio2018wild}. 
The application of deep learning to the cybersecurity domain does not constitute an exception to that. Recent classifiers proposed for malware detection, including the case of PDF, Android and malware binaries, have been indeed shown to be easily fooled by well-crafted adversarial manipulations~\cite{biggio2013evasion,demontis17-tdsc,grosse2016adversarial,kolosnjaji2018adversarial,kreuk2018adversarial,srndic14}.
Despite the sheer number of new malware specimen unleashed on the Internet (more than 8 millions in 2017 according to GData\footnote{\url{https://www.gdatasoftware.com/blog/2018/03/30610-malware-number-2017}}) demands for the application of effective automated techniques, the problem of adversarial examples has significantly questioned the suitability of deep-learning algorithms for these tasks. Nevertheless, it is not yet clear why such algorithms are easily fooled also in the particular application domain of malware detection.

In this work, we take a first step towards understanding the behavior of deep-learning algorithms for malware detection. To this end, we argue that explainable machine-learning algorithms, originally developed to interpret the black-box decisions of deep neural networks~\cite{sundararajan2017axiomatic,lime,guo2018lemna,koh17-icml}, can help unveiling the main characteristics learned by such algorithms to discriminate between benign and malicious files. In particular, we rely upon an explainable technique known as \emph{feature attribution}~\cite{sundararajan2017axiomatic} to identify the most influential input features contributing to each decision.
We focus on a case study related to the detection of Windows Portable Executable (PE) malware files, using a recently-proposed convolutional neural network named \emph{MalConv}~\cite{raff2017malware}. This network is trained directly on the raw input bytes to discriminate between malicious and benign PE files, reporting good classification accuracy. Recently, concurrent work~\cite{kolosnjaji2018adversarial,kreuk2018adversarial} has shown that it can be easily evaded by adding some carefully-optimized padding bytes at the end of the file, \ie, by creating \emph{adversarial malware binaries}.
However, no explanation has been clearly given behind the surprising vulnerability of this deep network. To address this problem, in this paper we adopt the aforementioned feature-attribution technique to provide meaningful explanations of the classification of malware binaries. Our underlying idea is to extend feature attribution to aggregate information on the most relevant input features at a higher semantic level, in order to highlight the most important instructions and sections present in each classified file.

Our empirical findings show that  \emph{MalConv} learns to discriminate between benign and malware samples mostly based on the characteristics of the file header, i.e., almost ignoring the data and text sections, which instead are the ones where the malicious content is typically hidden. This means that also depending on the training data, \emph{MalConv} may learn a spurious correlation between the class labels and the way the file headers are formed for malware and benign files. 

To further demonstrate the risks associated to using deep learning ``as is'' for malware classification, we propose a novel attack algorithm that generates adversarial malware binaries by only changing few tens of bytes in the file header. 
With respect to the other state-of-the-art attack algorithms in~\cite{kolosnjaji2018adversarial,kreuk2018adversarial}, our attack does not require injecting any padding bytes at the end of the file (it \emph{modifies} the value of some existing bytes in the header), and it is much more efficient, as it requires manipulating much fewer bytes.

The structure of this paper is the following: in \refToSection{classifiers} we introduce how to solve the problem of malware detection using machine learning techniques, and we present the architecture of MalConv; (ii) we introduce the integrated gradient technique as an algorithm that extracts which features contributes most for the classification problem; (iii) we collect the result of the method mentioned above applied on MalConv, highlighting its weaknesses, and (iv) we show how an adversary may exploit this information to craft an attack against MalConv.

\section{Deep Learning for Malware Detection in Binary Files}
\label{sec:classifiers}
\label{subsec:malconv}

Raff et al.~\cite{raff2017malware} propose \textit{MalConv}, a deep learning model which discriminates programs based on their byte representation, without extracting any feature.
The intuition of this approach is based on spatial properties of binary programs: ($i$) code, and data may be mixed, and it is difficult to extract proper features; ($ii$) there is correlation between different portions of the input program; ($iii$) binaries may have different length, as they are strings of bytes.
Moreover, there is no clear metrics that can be imposed on bytes: each value represents an instruction or a piece of data, hence it is difficult to set a distance between them.

\begin{figure}[t]
    \centering
    \includegraphics[width=\textwidth]{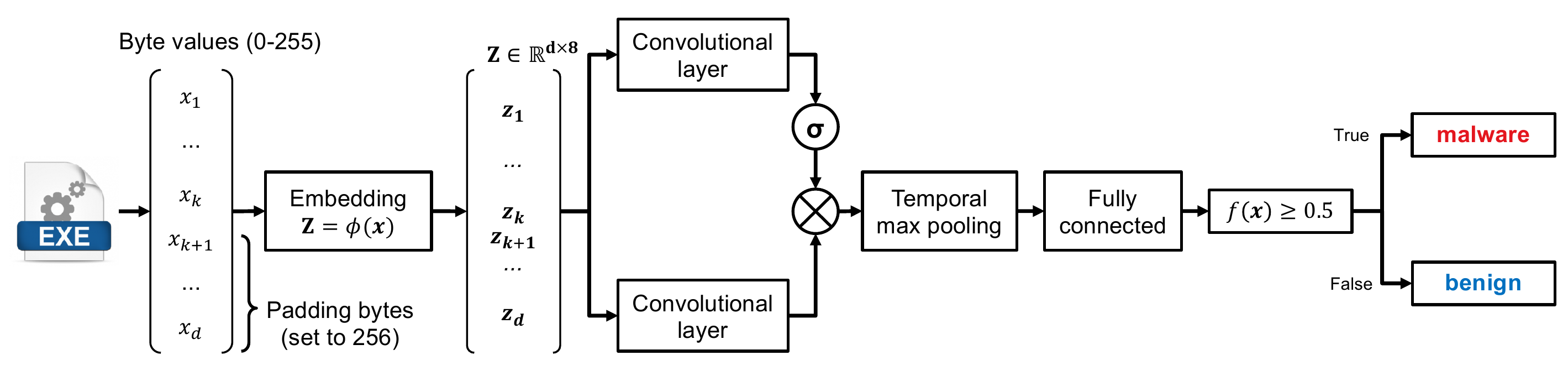}
    \caption{Architecture of MalConv~\cite{raff2017malware}, adapted from~\cite{kolosnjaji2018adversarial}.}
    \label{fig:malconv}
\end{figure}

To tackle all of these issues, Raff et al. develop a deep neural network that takes as input a whole program, whose architecture is represented in Fig.~\ref{fig:malconv}.
First, the input file size is bounded to $d=2^{21}$ bytes, i.e., $2$ MB.
Accordingly, the input file $\vct x$ is padded with the value $256$ if its size is smaller than $2$ MB; otherwise, it is cropped and only the first $2$ MB are analyzed. Notice that the padding value does not correspond to any valid byte, to properly represent the absence of information.
The first layer of the network is an embedding layer, defined by a function $\phi : \{0, 1, \ldots, 256\} \rightarrow \mathbb{R}^{d \times 8}$ that maps each input byte to an $8$-dimensional vector in the embedded space. This representation is learned during the training process and allows considering bytes as unordered categorical values.
The embedded sample $\mat Z$ is passed through two one-dimensional convolutional layers and then shrunk by a temporal max-pooling layer, before being passed through the final fully-connected layer.
The output $f(\vct x)$ is given by a softmax function applied to the results of the dense layer, and classifies the input $\vct x$ as malware if $f(\vct x)\geq 0.5$ (and as benign otherwise).

We debate the robustness of this approach, as it is not clear the rationale behind what has been learned by MalConv.
Moreover, we observe that the deep neural network focuses on the wrong sequences of bytes of an input binary.

\section{Explaining Machine Learning}
\label{sec:interpretation}
We address the problem mentioned at the end of Section~\ref{subsec:malconv} by introducing techniques aimed to explain the decisions of machine-learning models.
In particular, we will focus on techniques that explain the local predictions of such models, by highlighting the most influential features that contribute to each decision.
In this respect, linear models are easy to interpret: they rank each feature with a weight proportional to their relevance. Accordingly, it is trivial to understand which feature caused the classifier to assign a particular label to an input sample.
Deep-learning algorithms provide highly non-linear decision functions, where each feature may be correlated at some point with many others, making the interpretation of the result nearly impossible and leading the developers to naively trust their output.
Explaining predictions of deep-learning algorithms is still an open issue. Many researchers have proposed different techniques to explain what a model learns and which features mostly contribute to its decisions.
Among the proposed explanation methods for machine learning, we decided to focus on a technique called \textit{integrated gradients}~\cite{sundararajan2017axiomatic}, developed by Sundararajan et al., for two main reasons. First, it does not use any learning algorithm to explain the result of another machine learning model; and, second, it is more efficient w.r.t. the computations that are required by the other methods in the state-of-the-art.

\myparagraph{Integrated gradients} We introduce the concept of \textit{attribution methods}: algorithms that compute the contribution of each feature for deciding which label needs to be assigned to a single point.
Contributions are calculated w.r.t. a \textit{baseline}.
A baseline is a point in the input space that corresponds to a null signal: for most image classifiers it may be identified as a black image, it is an empty sentence for text recognition algorithms, and so on.
The baseline serves as ground truth for the model: each perturbation to the baseline should increase the contributions computed for the modified features.
Hence, each contribution is computed w.r.t. the output of the model on the baseline.
The integrated gradients technique is based on two axioms and upon the concept of baseline.

\myparagraph{Axiom I: Sensitivity}

The first axiom is called \textit{sensitivity}: an attribution method satisfies sensitivity if, for every input that differ in one feature from the baseline but they are classified differently, then the attribution of the differing feature should be non-zero.

Moreover, if the learned function does not mathematically depend on a particular feature, the attribution should be zero.
On the contrary, if sensitivity is not satisfied, the model is focusing on irrelevant features, as the attribution method fails to weight the contribution of each variable.
Authors state that gradients violate the sensitivity axioms, hence using them during the training phase by applying back-propagation implies attributing wrong importance to the wrong features.

\myparagraph{Axiom II: Implementation Invariance}
The second axiom is called \textit{implementation invariance}, and it is built on top of the notion of \textit{functional equivalence}: two networks are functionally equivalent if their outputs are equal on all inputs, despite being implemented in different ways.
Thus, an attribution method satisfies implementation invariance if it produces the same attributions for two functionally equivalent networks.
On the contrary, if this axiom is not satisfied, the attribution method is sensitive to the presence of useless aspects of the model.
On top of these two axioms, Sundararajan et al. propose the integrated gradient method that satisfies both sensitivity and implementation invariance.
Hence, this algorithm should highlight the properties of the input model and successfully attributing the correct weights to the feature believed relevant by the model itself.

\myparagraph{Integrated Gradients}
Given the input model $f$, a point $\vct x$ and baseline $\vct x^\prime$, the attribution for the $i^{\rm th}$ feature is computed as follows:
\begin{equation}
    \begin{split}
        IG_i(\vct x) = (x_i - x_i') \int_0^1 \frac{\partial f(\vct x' + \alpha( \vct x - \vct x'))}{\partial x_i}d\alpha
    \end{split} \, .
    \label{eq:int0}
\end{equation}
This is the integral of the gradient computed on all points that lie on the line passing through $\vct x$ and $\vct x'$.
If $\vct x'$ is a good baseline, each point in the line should add a small contribution to the classification output.
This method satisfies also the \textit{completeness} axiom: the attributions add up to the difference between the output of the model at the input $\vct x$ and $\vct x'$. 
Hence, the features that are important for the classification of $\vct x$ should appear by moving on that line.

Since we can only compute discrete quantities, the integral can be approximated using a summation, adding a new degree of freedom to the algorithm, that is the number of points to use in the process: Sundararajan et al. state that the number of steps could be chosen between 20 and 300, as they are enough to approximate the integral within the 5\% of accuracy.

\section{What Does MalConv Learn?}
\label{sec:what_learns}
We applied the integrated gradient technique for trying to grasp the intuition of what is going on under the hood of MalConv deep network.
For our experiments we used a simplified version of MalConv, with an input dimension shrunk to $2^{20}$ instead of $2^{21}$, that is 1 MB instead of 2 MB, trained by Anderson et al.~\cite{2018arXiv180404637A} and publicly available on GitHub~\footnote{\url{https://github.com/endgameinc/ember/tree/master/malconv}}.
To properly comment the result generated by the attribution method, we need to introduce the layout of the executables that run inside Windows operating system.

\myparagraph{Windows Portable Executable format}
The Windows Portable Executable format\footnote{\url{https://docs.microsoft.com/en-us/windows/desktop/debug/pe-format}} (PE) describes the structure of a Windows executable.
Each program begins with the DOS header, which is left for retro-compatibility issues and for telling the user that the program can not be run in DOS.
The only two useful information contained into the DOS header are the DOS magic number \texttt{MZ} and the value contained at offset \texttt{0x3c}, that is an offset value that point to the real PE header.
If the first one is modified, Windows throws an error and the program is not loaded, while if the second is perturbed, the operating system can not find the metadata of the program.
The PE header consists of a Common Object File Format (COFF) header, containing metadata that describe which Windows Subsystem can run that file, and more.
If the file is not a library, but a standalone executable, it is provided with an Optional Header, which contains information that are used by the loader for running the 
executable. As part of the optional header, there are the section headers.
The latter is composed by one entry for each section in the program.
Each entry contains meta-data used by the operating system that describes how that particular section needs to be loaded in memory.
There are other components specified by the Windows PE format, but for this work this information is enough to understand the output of the integrated gradients algorithm applied to MalConv.

\myparagraph{Integrated gradients applied to malware programs} The integrated gradients technique works well with image and text files, but we need to evaluate it on binary programs.
First of all, we have to set a baseline: since the choice of the latter is crucial for the method to return accurate results, we need to pick a sample that satisfies the constraints\footnote{\url{https://github.com/ankurtaly/Integrated-Gradients/blob/master/howto.md}} highlighted by the authors of the method: (i) the baseline should be an empty signal with null response from the network; (ii) the entropy of the baseline should be very low. 
If not, that point could be an adversarial point for the network and not suitable for this method.

Hence, we have two possible choices: (i) the empty file, and (ii) a file filled with zero bytes.
For MalConv, an empty file is a vector filled by the special padding number $256$, as already mentioned in Section~\ref{subsec:malconv}.
While both these baselines satisfy the second constraint, only the empty file has a null response, as the zero vector is seen as malware with the 20\% of confidence.
Hence, the empty file is more suitable for being considered the ground truth for the network.
The method returns a matrix that contains all the attributions, feature per feature, $\mat V \in \mathbb{R}^{d \times 8}$, where the second dimension is produced by the embedding layer.
We compute the mean for each row of $\mat V$ to obtain a signed scalar number for each feature, resulting in a point $\vct v' \in \mathbb{R}^{d}$ and it may be visualized in a plot.
\begin{figure}
    \centering
    \includegraphics[width=\textwidth]{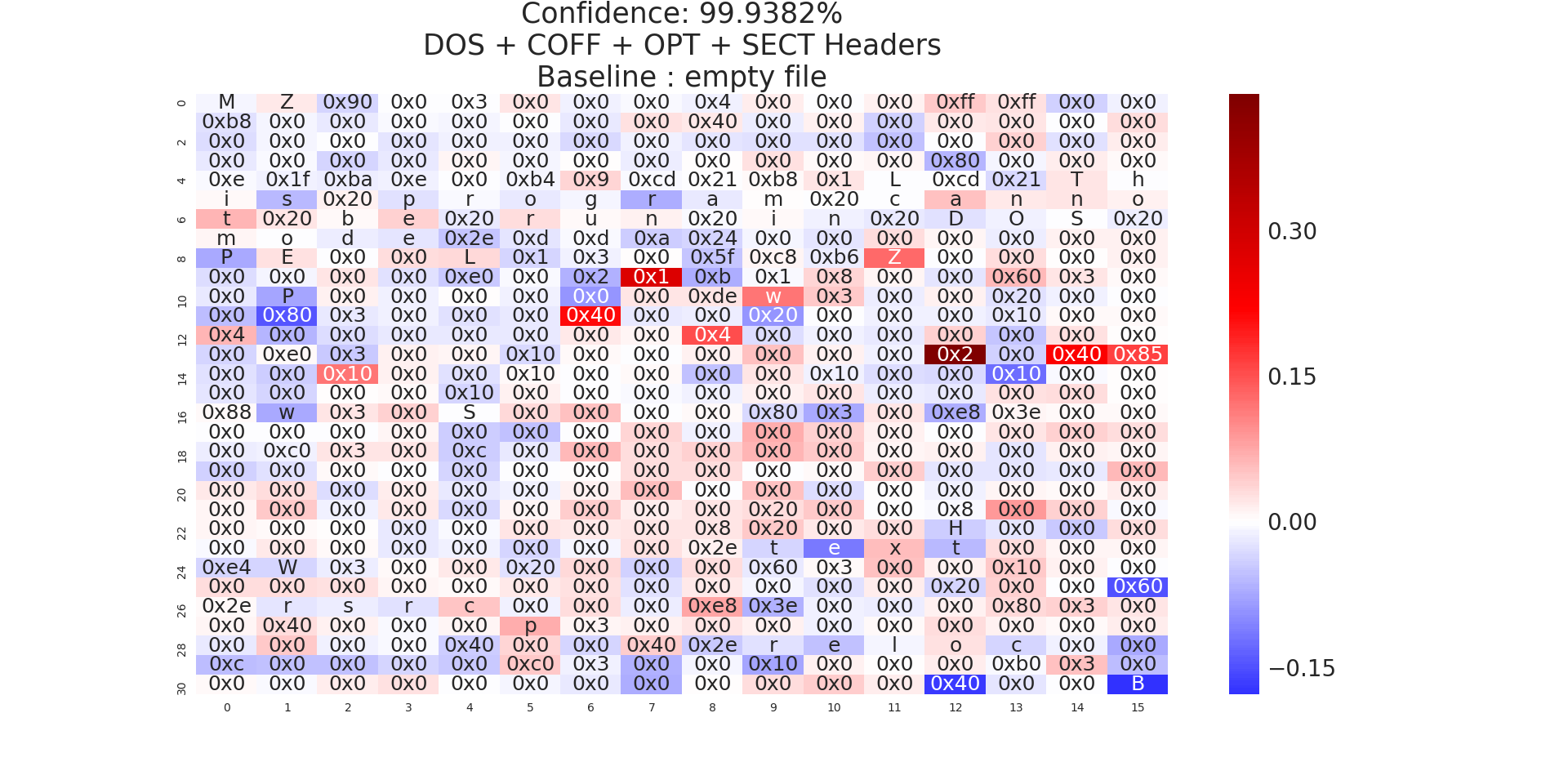}
    \caption{Attribution given by MalConv to the header of a malware sample.}
    \label{fig:header_malw0}
\end{figure}
Figure~\ref{fig:header_malw0} highlights the importance attributed by MalConv to the header of an input malware example using the empty file as baseline, marking with colours each byte's contribution. 
We can safely state that a sample is malware if it is scored as such by VirusTotal\footnote{\url{https://www.virustotal.com}} which is an online service for scanning suspicious files.
The cells colored in red symbolize that MalConv considered them for deciding whether the input sample is malware.
On the contrary, cells with blue background are the ones that MalConv retains representative of benign programs.
Regarding our analysis, we may notice that MalConv is giving importance to portions of the input program that are not related to any malicious intent: some bytes in the DOS header are colored both in red and blue, and a similar situation is found for the other sections.
Notably, the DOS header is not even read by modern operating systems, as the only important values are the magic number and the value contained at offset \texttt{0x3c}, as said in the previous paragraph.
All the other bytes are repeated in every Windows executable, and they are not peculiar neither for goodware nor malware programs.
We can thus say that this portion of the program should not be relevant for discriminating benign from malicious programs. Accordingly, one would expected MalConv to give no attribution to these portions of bytes.

\begin{figure}
    \centering
    \includegraphics[height=0.3\textheight]{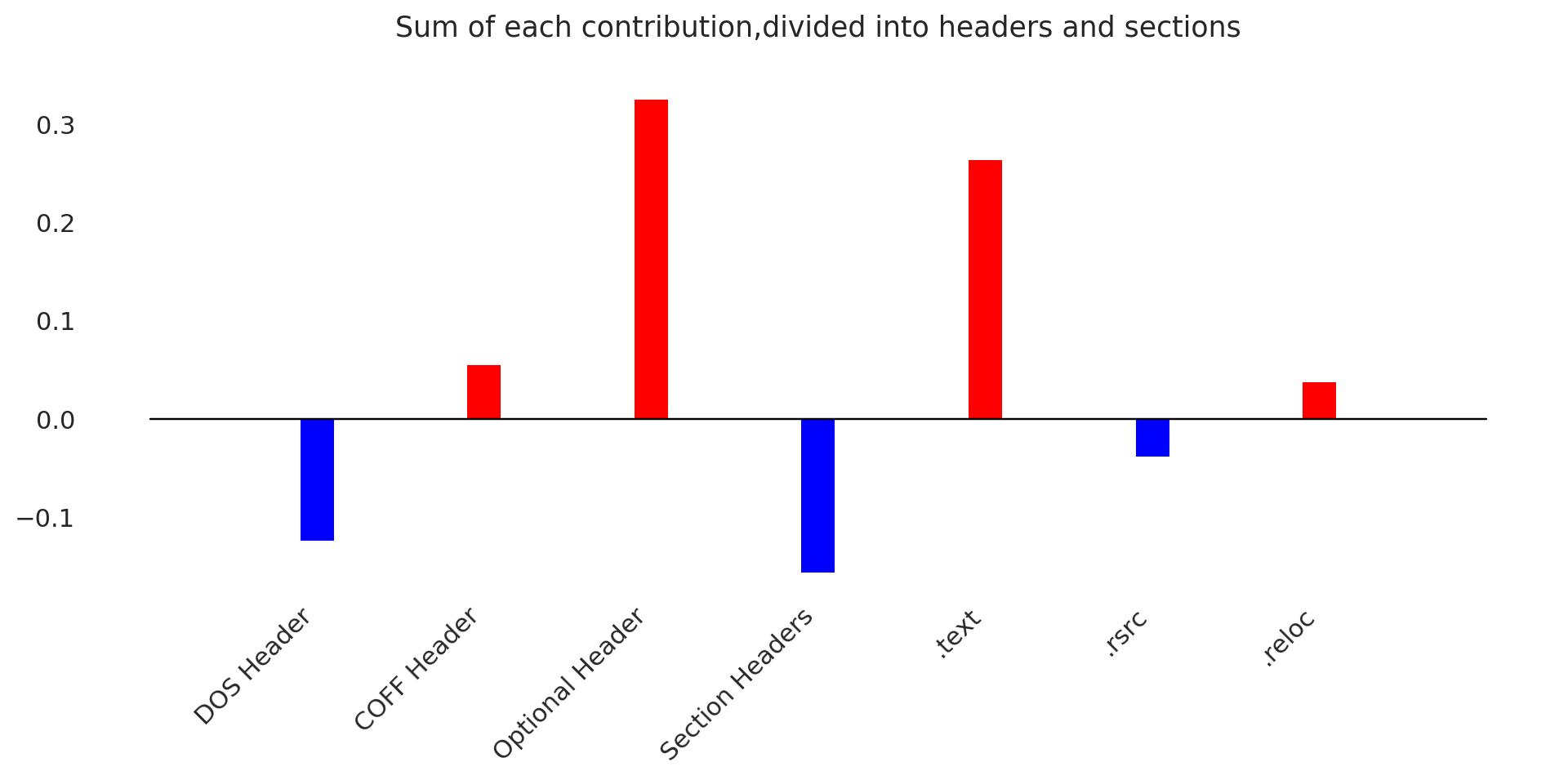}
    \caption{Sum of contributions expressed in percentage.}
    \label{fig:aggr_contributions}
\end{figure}

We can observe the attribution assigned by integrated gradients to every byte of the program, aggregated for each component of the binary, in Figure~\ref{fig:aggr_contributions}.
Each entry of the histogram is the sum of all the contributions given by each byte in that region.
The histogram is normalized using the $l1$ norm of its components.
The color scheme is the same as the one described for Figure~\ref{fig:header_malw0}.
MalConv puts higher weights at the beginning of the file, and this fact has been already formalized by Kolosnjaji et al.~\cite{kolosnjaji2018adversarial} in the discussion of the results of the attack they have developed.
It is clear that the \textit{.text} section has a large impact on the classification, as it is likely that the maliciousness of the input program lies in that portion of the malware.
However, the contribution given by the COFF and optional headers outmatch the other sections. 
This further implies that the locations that are learned as important for classification are somewhat misleading.
We can state that among all the correlations that are present inside a program, MalConv surely learned something that is not properly relevant for classification of malware and benign programs.
By knowing this fact, an adversary may perturb these portions of malware and sneaking behind MalConv without particular effort: all she or he has to do is to compute the gradient of the model w.r.t. to the input.
Hence, we describe how a malicious user can deliver an attack by perturbing bytes contained into an input binary to evade the classifier.

\section{Evading Malconv by Manipulating the File Header}
\label{sec:attack}
We showed that MalConv bases its classifications upon unreliable and spurious features, as also indirectly shown by state-of-the-art attacks against the same network, which inject bytes in specific regions of the malware sample without even altering the embedded malicious code~\cite{kolosnjaji2018adversarial,kreuk2018adversarial}.
In fact, these attacks can not alter the semantics of the malware sample, as it needs to evade the classifier while preserving its malicious functionality. Hence, not all the bytes in the input binary can be modified, as it is easy to alter a single instruction and break the behavior of the original program.
As already mentioned in Section~\ref{sec:intro}, the only bytes that are allowed to be perturbed are those placed in unused zones of the program, such as the space that is left by the compiler between sections, or at the end of the sample.
Even though these attacks have been proven to be effective, we believe that it is not necessary to deploy such techniques as the network under analysis is not learning what the developers could have guessed during training and test phases.
Recalling the results shown in Section~\ref{sec:what_learns}, we believe that, by perturbing only the bytes inside the DOS header, malware programs can evade MalConv.
There are two constraints: the \texttt{MZ} magic number and the value at offset $0x3c$ can not be modified, as said in Section~\ref{sec:what_learns}.
Thus, we flag as editable all the bytes inside the DOS header that are contained in that range.
Of course, one may also manipulate more bytes, but we just want to highlight the severe weaknesses anticipated in Section~\ref{sec:interpretation}.

\myparagraph{Attack Algorithm} Our attack algorithm is given as Algorithm~\ref{alg:algo1}. 
It first computes the representation of each byte of the input file $\vct x$ in the embedded space, as $\mat Z \leftarrow \phi(\vct x)$.
The gradient $\mat G$ of the classification function $f(\vct x)$ is then computed w.r.t. the embedding layer. 
Note that we denote with $\vct g_i, \vct z_i \in \mathbb R^8$ the $i^{\rm th}$ row of the aforementioned matrices. 
For each byte indexed in $I$, \ie, in the subset of header bytes that can be manipulated, our algorithm follows the strategy implemented by Kolosnjaji et al.~\cite{kolosnjaji2018adversarial}. 
This amounts to selecting the closest byte in the embedded space which is expected to maximally increase the probability of evasion. 
The algorithm stops either if evasion is achieved or if it reaches a maximum number of allowed iterations.
The latter is imposed since, depending on which bytes are allowed to be modified, evasion is not guaranteed.

The result of the attack applied using an example can be appreciated in Figure~\ref{fig:evading}: the plot shows the probability of being recognized as malware, which is the curve coloured in blue.
Each iteration of the algorithm is visualized as black vertical dashed lines, showing that we need to interpret the gradient different times to produce an adversarial example.
This is only one example of the success of the attack formalized above, as we tested it on 60 different malware inputs, taken from both \textit{The Zoo}\footnote{\url{https://github.com/ytisf/theZoo}} and \textit{Das Malverik}.\footnote{\url{http://dasmalwerk.eu/}}
Experimental results confirm our doubts: 52 malware programs over 60 evade MalConv by only perturbing the DOS header.
Note that a human expert would not be fooled by a similar attack, as the DOS header can be easily stripped and replaced with a normal one without perturbed bytes. 
Accordingly, \emph{MalConv} should have learned that these bytes are not relevant to the classification task.

\begin{minipage}[c]{\textwidth}
    \begin{minipage}{0.55\textwidth}
        \begin{algorithm}[H]
            \SetKwInOut{Input}{input}
            \SetKwInOut{Given}{given}
            \SetKwInOut{Output}{output}
            \caption{Evasion algorithm}
            \label{alg:algo1}
            \SetAlgoLined
            \Input{$\vct x$ input binary, $I$ indexes of bytes to perturb, $f$ classifier, $T$ max iterations.}
            \Output{Perturbed code that should achieve evasion against $f$.}
            \BlankLine
            $t \leftarrow 0$\;
            \While{ $f(\vct x) \geq 0.5 \wedge t < T$ }{
                $\mat Z \leftarrow \phi(\vct x)\in \mathbb{R}^{d \times 8}$\;
                $\mat G \leftarrow -\nabla_{\phi} f(\vct x) \in \mathbb{R}^{d \times 8}$\;
                \ForAll{$i \in I$}{
                    $\vct g_i \leftarrow \vct g_i / \| \vct g_i \|_2$\;
                    $\vct s \leftarrow \vct 0 \in \mathbb{R}^{256}$\;
                    $\vct d \leftarrow \vct 0 \in \mathbb{R}^{256}$\;
                    \ForAll{$b \in 0,...,255$}{
                        $\vct z_b \leftarrow \phi(b)$\;
                        $s_b \leftarrow \vct g_i^\T \cdot (\vct z_b - \vct z_i)$\;
                        $d_b \leftarrow \| \vct z_b - (\vct z_i + s_b \cdot \vct g_i) \|$\;
                    }
                    $x_i \leftarrow \argmin_{j:s_j > 0} d_j$\;
                }
                $t \leftarrow t+1$\;
            }
        \end{algorithm}
    \end{minipage}
    \begin{minipage}{0.45\textwidth}
        \centering
        \includegraphics[width=\textwidth]{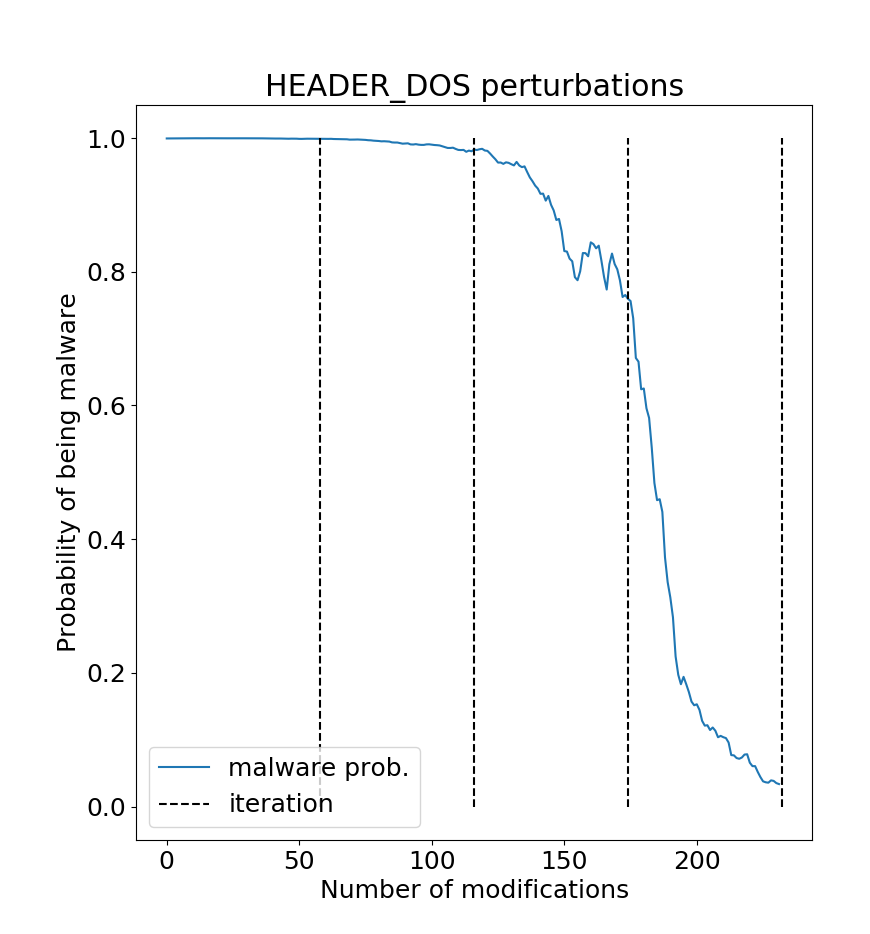}
        \captionof{figure}{Evading MalConv by perturbing only few bytes in the DOS header. 
        The black dashed lines represent the iterations of the algorithm. In each iteration, the algorithm manipulates $58$ header bytes (excluding those that can not be changed). 
        The blue curve shows the probability of the perturbed sample being classified as malware across the different iterations.}
        \label{fig:evading}
    \end{minipage}
\end{minipage}

\section{Related Work}

We discuss here some approaches that leverage deep learning for malware detection, which we believe may exhibit similar problems to those we have highlighted for \emph{MalConv} in this work.
We continue with a brief overview of the already proposed attacks that address vulnerabilities in the MalCon architecture.
Then, we discuss other explainable machine-learning techniques that may help us to gain further insights on what such classifiers effectively learn, and potentially whether and to which extent we can make them more reliable, transparent and secure.

\myparagraph{Deep Malware Classifiers} Saxe and Berlin~\cite{saxe2015deep} use a deep learning system trained on top of features extracted from Windows executables, such as byte entropy histograms, imported functions, and meta-data harvested from the header of the input executable.
Hardy et al.~\cite{hardy2016dl4md} propose DL4MD, a deep neural network for recognizing Windows malware programs.
Each sample is characterized by its API calls, hence they are passed to a stacked auto-encoders network.
David et al.~\cite{david2015deepsign} use a similar approach, by exploiting de-noising auto-encoders for classifying Windows malware programs.
The features used by the author are harvested from dynamic analysis.
Yhuan et al.~\cite{yuan2016droiddetector} extract features from both static and dynamic analysis of an Android application, like the requested permissions, API calls, network usage, sent SMS and so on, and samples are given in input to a deep neural network.
McLaughlin et al.~\cite{mclaughlin2017deep} propose a convolutional neural network which takes as input the sequence of the opcodes of an input program.
Hence, they do not extract any features from data, but they merely extract the sequence of instructions to feed to the model.

\myparagraph{Evasion attacks against MalConv} Kolosnjaji et al.~\cite{kolosnjaji2018adversarial} propose an attack that does not alter bytes that correspond to code, but they append bytes at the end of the sample, preserving its semantics.
The padding bytes are calculated using the gradient of the cost function w.r.t. the input byte in a specific location in the malware, achieving evasion with high probability.
The technique is similar to the one developed by Goodfellow et al.~\cite{goodfellow6572explaining}, the so-called Fast Sign Gradient Method (FSGM).
Similarly, Kreuk et al.~\cite{kreuk2018adversarial} propose to append bytes at the end of the malware and fill unused bytes between sections. %
Their algorithm is different: they work in the embedding domain, and they translate back in byte domain after applying the FSGM to the malware, while Kolosnjaji et al. modify one padding byte at the time in the embedded domain, and then they translate it back to the byte domain.

\myparagraph{Explainable Machine Learning} Riberio et al.~\cite{lime} propose a method called LIME, which is an algorithm that tries to explain which features are important w.r.t. the classification result.
It takes as input a model, and it learns how it behaves locally around an input point, by producing a set of artificial samples.
The algorithm uses LASSO to select the top K features, where K is a free parameter.
As it may be easily noticed, the dimension of the problem matters, and it may become computational expensive on high dimensional data.
Guo et al.~\cite{guo2018lemna} propose LEMNA, which is an algorithm specialized on explaining deep learning results for security applications.
As said in Section~\ref{sec:classifiers}, many malware detectors exploit machine learning techniques in their pipeline.
Hence, being able to interpret the results of the classification may be helpful to analysts for identifying vulnerabilities.
Guo et al. use a Fused LASSO or a mixture Regression Model to learn the decision boundary around the input point and compute the attribution for $K$ features, where $K$ is a free parameter.
Another important work that may help us explaining predictions (and misclassifications) of deep-learning algorithms for malware detection is that by Koh and Liang~\cite{koh17-icml}, which allows identifying the most relevant training prototypes supporting classification of a given test sample. With these technique, it may be possible to associate or compare a given test sample to some known training malware samples and their corresponding families (or, to some relevant benign file).
\section{Conclusions and Future Work}

We have shown that, despite the success of deep learning in many areas, there are still uncertainties regarding the precision of the output of these techniques.
In particular, we use \textit{integrated gradients} to explain the results achieved by MalConv, highlight the fact that the deep neural network attributes wrong non-zero weights to well known useless features, that in this case are locations inside a Windows binary.

To further explain these weaknesses, we devise an algorithm inspired by recent papers in adversarial machine learning, applied to the malware detection.
We show that perturbing few bytes is enough for evading MalConv with high probability, as we applied this technique on a concrete set of samples taken from the internet, achieving evasion on most of all the inputs. 

Aware of this situation, we are willing to keep investigating the weaknesses of these deep learning malware classifiers, as they are becoming more and more important in the cybersecurity world.
In particular, we want to devise attacks that can not be easily recognized by a human expert: the perturbations in the DOS header are easy to detect, as mentioned in Section~\ref{sec:attack} and it may be patched without substantial effort.
Instead, we are studying how to hide these modifications in an unrecognizable way, like filling bytes between functions: compilers often leave space between one function and the other to align the function entry point to multiple of $32$ or $64$, depending on the underlying architecture.
The latter is an example we are currently investigating.
Hence, all classifiers that rely on raw byte features may be vulnerable to such attacks, and our research may serve as a proposal for a preliminary penetration testing suite of attacks that developers could use for establishing the security of these new technologies.

In conclusion, we want to emphasize that these intelligent technologies are far from being considered secure, as they can be easily attacked by aware adversaries.
We hope that our study may focus the attention on this problem and increase the awareness of developers and researchers about the vulnerability of these models.
\bibliographystyle{unsrt}
\bibliography{reference,bibDB}

\begin{thebibliography}{10}

\bibitem{biggio2013evasion}
Battista Biggio, Igino Corona, Davide Maiorca, Blaine Nelson, Nedim
  {\v{S}}rndi{\'c}, Pavel Laskov, Giorgio Giacinto, and Fabio Roli.
\newblock Evasion attacks against machine learning at test time.
\newblock In {\em Joint European conference on machine learning and knowledge
  discovery in databases}, pages 387--402. Springer, 2013.

\bibitem{szegedy14-iclr}
Christian Szegedy, Wojciech Zaremba, Ilya Sutskever, Joan Bruna, Dumitru Erhan,
  Ian Goodfellow, and Rob Fergus.
\newblock Intriguing properties of neural networks.
\newblock In {\em International Conference on Learning Representations}, 2014.

\bibitem{goodfellow6572explaining}
Ian~J Goodfellow, Jonathon Shlens, and Christian Szegedy.
\newblock Explaining and harnessing adversarial examples (2014).
\newblock {\em arXiv preprint arXiv:1412.6572}.

\bibitem{papernot2016limitations}
Nicolas Papernot, Patrick McDaniel, Somesh Jha, Matt Fredrikson, Z~Berkay
  Celik, and Ananthram Swami.
\newblock The limitations of deep learning in adversarial settings.
\newblock In {\em Security and Privacy (EuroS\&P), 2016 IEEE European Symposium
  on}, pages 372--387. IEEE, 2016.

\bibitem{carlini2017towards}
Nicholas Carlini and David Wagner.
\newblock Towards evaluating the robustness of neural networks.
\newblock In {\em 2017 IEEE Symposium on Security and Privacy (SP)}, pages
  39--57. IEEE, 2017.

\bibitem{biggio2018wild}
Battista Biggio and Fabio Roli.
\newblock Wild patterns: Ten years after the rise of adversarial machine
  learning.
\newblock {\em Pattern Recognition}, 84:317--331, 2018.

\bibitem{demontis17-tdsc}
Ambra Demontis, Marco Melis, Battista Biggio, Davide Maiorca, Daniel Arp,
  Konrad Rieck, Igino Corona, Giorgio Giacinto, and Fabio Roli.
\newblock Yes, machine learning can be more secure! a case study on android
  malware detection.
\newblock {\em IEEE Trans. Dependable and Secure Computing}, In press.

\bibitem{grosse2016adversarial}
Kathrin Grosse, Nicolas Papernot, Praveen Manoharan, Michael Backes, and
  Patrick McDaniel.
\newblock Adversarial perturbations against deep neural networks for malware
  classification.
\newblock {\em arXiv preprint arXiv:1606.04435}, 2016.

\bibitem{kolosnjaji2018adversarial}
Bojan Kolosnjaji, Ambra Demontis, Battista Biggio, Davide Maiorca, Giorgio
  Giacinto, Claudia Eckert, and Fabio Roli.
\newblock Adversarial malware binaries: Evading deep learning for malware
  detection in executables.
\newblock {\em arXiv preprint arXiv:1803.04173}, 2018.

\bibitem{kreuk2018adversarial}
Felix Kreuk, Assi Barak, Shir Aviv-Reuven, Moran Baruch, Benny Pinkas, and
  Joseph Keshet.
\newblock Adversarial examples on discrete sequences for beating whole-binary
  malware detection.
\newblock {\em arXiv preprint arXiv:1802.04528}, 2018.

\bibitem{srndic14}
Nedim \v{S}rndic and Pavel Laskov.
\newblock Practical evasion of a learning-based classifier: A case study.
\newblock In {\em Proc. 2014 IEEE Symp. Security and Privacy}, SP '14, pages
  197--211, Washington, DC, USA, 2014. IEEE CS.

\bibitem{sundararajan2017axiomatic}
Mukund Sundararajan, Ankur Taly, and Qiqi Yan.
\newblock Axiomatic attribution for deep networks.
\newblock {\em arXiv preprint arXiv:1703.01365}, 2017.

\bibitem{lime}
Marco~Tulio Ribeiro, Sameer Singh, and Carlos Guestrin.
\newblock "why should {I} trust you?": Explaining the predictions of any
  classifier.
\newblock In {\em Proceedings of the 22nd {ACM} {SIGKDD} International
  Conference on Knowledge Discovery and Data Mining, San Francisco, CA, USA,
  August 13-17, 2016}, pages 1135--1144, 2016.

\bibitem{guo2018lemna}
Wenbo Guo, Dongliang Mu, Jun Xu, Purui Su, Gang Wang, and Xinyu Xing.
\newblock Lemna: Explaining deep learning based security applications.
\newblock In {\em Proceedings of the 2018 ACM SIGSAC Conference on Computer and
  Communications Security}, pages 364--379. ACM, 2018.

\bibitem{koh17-icml}
P.~W. Koh and P.~Liang.
\newblock Understanding black-box predictions via influence functions.
\newblock In {\em International Conference on Machine Learning (ICML)}, 2017.

\bibitem{raff2017malware}
Edward Raff, Jon Barker, Jared Sylvester, Robert Brandon, Bryan Catanzaro, and
  Charles Nicholas.
\newblock Malware detection by eating a whole exe.
\newblock {\em arXiv preprint arXiv:1710.09435}, 2017.

\bibitem{2018arXiv180404637A}
H.~S. {Anderson} and P.~{Roth}.
\newblock {EMBER: An Open Dataset for Training Static PE Malware Machine
  Learning Models}.
\newblock {\em ArXiv e-prints}, April 2018.

\bibitem{saxe2015deep}
Joshua Saxe and Konstantin Berlin.
\newblock Deep neural network based malware detection using two dimensional
  binary program features.
\newblock In {\em Malicious and Unwanted Software (MALWARE), 2015 10th
  International Conference on}, pages 11--20. IEEE, 2015.

\bibitem{hardy2016dl4md}
William Hardy, Lingwei Chen, Shifu Hou, Yanfang Ye, and Xin Li.
\newblock Dl4md: A deep learning framework for intelligent malware detection.
\newblock In {\em Proceedings of the International Conference on Data Mining
  (DMIN)}, page~61. The Steering Committee of The World Congress in Computer
  Science, Computer Engineering and Applied Computing (WorldComp), 2016.

\bibitem{david2015deepsign}
Omid~E David and Nathan~S Netanyahu.
\newblock Deepsign: Deep learning for automatic malware signature generation
  and classification.
\newblock In {\em Neural Networks (IJCNN), 2015 International Joint Conference
  on}, pages 1--8. IEEE, 2015.

\bibitem{yuan2016droiddetector}
Zhenlong Yuan, Yongqiang Lu, and Yibo Xue.
\newblock Droiddetector: android malware characterization and detection using
  deep learning.
\newblock {\em Tsinghua Science and Technology}, 21(1):114--123, 2016.

\bibitem{mclaughlin2017deep}
Niall McLaughlin, Jesus Martinez~del Rincon, BooJoong Kang, Suleiman Yerima,
  Paul Miller, Sakir Sezer, Yeganeh Safaei, Erik Trickel, Ziming Zhao, Adam
  Doupe, et~al.
\newblock Deep android malware detection.
\newblock In {\em Proceedings of the Seventh ACM on Conference on Data and
  Application Security and Privacy}, pages 301--308. ACM, 2017.

\end{thebibliography}
\end{document}